\documentclass[preprint]{aastex}

\shorttitle{Southern close binary stars.~I}
\shortauthors{Duerbeck \& Rucinski}

\begin{document}

\title{Radial Velocity Studies of Southern Close Binary 
Stars.~I\footnote{Based on the data obtained at the European 
Southern Observatory.}}

\author{Slavek M. Rucinski}
\affil{David Dunlap Observatory, University of Toronto \\
P.O.~Box 360, Richmond Hill, Ontario, Canada L4C~4Y6}
\email{rucinski@astro.utoronto.ca}

\and

\author{Hilmar Duerbeck}
\affil{WE/OBSS, Vrije Universiteit Brussel, Pleinlaan 2, 
B-1150 Brussels, Belgium}
\email{hduerbec@vub.ac.be}

\begin{abstract}
Radial-velocity measurements and sine-curve fits to the orbital 
velocity variations are presented for nine contact binaries,
V1464~Aql, V759~Cen, DE~Oct, MW~Pav, BQ~Phe, 
EL~Aqr, SX~Crv, VZ~Lib, GR~Vir; for the first five among these,
our observations are the first available radial velocity data.
Among three remaining radial velocity variables, 
CE~Hyi is a known visual binary, while CL~Cet and V1084~Sco
are suspected to be multiple systems where the
contact binary is spectrally dominated by its companion (which
itself is a binary in V1084~Sco). Five additional variables, 
V872 Ara, BD~Cap, HIP~69300, BX~Ind, V388~Pav,
are of unknown type, but most are pulsating stars; 
we give their mean radial velocities and $V \sin i$.
\end{abstract}


\section{INTRODUCTION}
\label{sec1}

The origins of this paper are related to those of the  
series of radial velocity studies of short period binaries 
currently conducted at the David Dunlap Observatory 
(DDO papers 1 -- 10) 
\citep{ddo1,ddo2,ddo3,ddo4,ddo5,ddo6,ddo8,ddo9,ddo10}. 
Both authors realized in the 1990's that with availability
of good Hipparcos parallaxes \citep{hip},
the limiting factor in gathering spatial velocities of
contact binaries would be radial velocities (RV). While the DDO
studies have since succeeded in obtaining RV data for now
over one hundred of northern binaries, for many reasons the data 
presented in this paper are so far the only effort for the
southern binaries. We present these results because chances
of continuation of these observations is basically nil:
the telescope has been retired, and the remaining ESO telescopes 
are assigned for technically more demanding tasks.

The observations reported in this paper have been
collected on 4 nights of August 8 -- 11, 1998.
To optimize on the returns from such a short survey, 
the 17 targets were selected to be a mixture
of contact binaries possibly offering reasonable orbital
solutions with a selection of variables suspected to be
contact binaries \citep{Duer97}. The next paper will contain 
similar results for Spring southern targets.

In this paper, we attempt to stay close to the format of the
DDO series. In particular, we use the same 
data extraction procedures through the Broadening Function (BF)
approach, as described in the DDO interim summary paper 
\citet[DDO-7]{ddo7}. Of 17 stars discussed in this
paper, four contact binaries (EL~Aqr, SX~Crv, VZ~Lib, GR~Vir) 
in the meantime have been observed during the DDO program resulting in 
good RV orbits; we include these systems here to report
the Southern observations as a check of consistency.
The remaining stars have been observed by us
for radial-velocity variations for the first time.
We have derived the radial velocities in the same way as described
in the DDO papers; see the DDO-7 paper for a discussion of the
broadening-function technique used in the derivation of
the radial-velocity orbit parameters:
the amplitudes, $K_i$, the center-of-mass
velocity, $V_0$, and the time-of-primary-eclipse epoch, $T_0$.
The primary radial velocity standard used to determine the
BF's as well as to find radial velocities was 6~Cet (F5V)
assumed to have the velocity of +14.9 km~s$^{-1}$ (Simbad).
This was the only sufficiently well observed standard
which could be used as the BF template, but appeared 
to serve well for the whole range of the spectral 
types from mid-A to mid-G; the
disparity of the spectral types manifested itself mostly
in the broadening function intensities which 
would not normalize to unity as expected for perfect
spectral matches. 

We describe our results in the context of the existing
photometric data from the literature and the Hipparcos
project. We also utilize the mean $(B-V)$ color indexes 
taken from the Tycho-2 catalog \citep{Tycho2}
and the photometric estimates of the spectral types
using the relations published by \citet{Bessell1979}.
The spectral types are taken uniformly from the 5 volumes
of the Michigan Catalogue of HD Stars  \citep{houk}; 
from now on called HDH.
Because the high incidence of companions to contact binary stars
\citep{PR2006}, we checked all stars for possible 
membership in visual systems using the Washington Double Star 
Catalog (WDS)\footnote{http://ad.usno.navy.mil/wds/}.
DE~Oct and CE~Hyi have been identified as members of
already known visual binaries. VZ~Lib is a previously recognized
\citep[DDO-4]{ddo4} spectroscopic triple system.

The observations were carried with the ESO 1.52-m telescope
at ESO La Silla, equipped with a Boller \& Chivens Cassegrain
spectrograph. Holographic grating No.~32 (2400 lines/mm) was 
used in combination with Loral CCD No.~39 ($2048\times 2048$ pix). 
The slit with was set to 220 $\mu$m.
The broadening functions were extracted from the 
wavelength region of  401.6 -- 499.8 nm. Thus, compared with
the DDO results based on the Mg~I triplet at 518.4 nm,
with the window of about 25 -- 30 nm, the ESO spectra are longer
and more blue. They also have a lower resolution:
While for the DDO spectra, the broadening functions have
a resolution of typically $\sigma \simeq 13 - 18$ km~s$^{-1}$,
the spectra described here have the BF resolution of 
typically $\sigma \simeq 23 - 27$ km~s$^{-1}$.
Stellar exposure times ranged between 10 and 20 min, depending on the
brightness of the object; each exposure was followed by the exposure
of a He-Ar spectrum. Spectrum extraction and wavelength calibration was
carried out using the ESO MIDAS software 
system\footnote{http://www.eso.org/projects/esomidas}. 

This paper is structured in a way similar to the DDO series. 
We comment on individual contact binary systems
in Section~\ref{sec2}. 
Three additional variables which may be members of multiple
systems with a dominating third component are described in
Section~\ref{sec3}, while additional five stars of mostly unknown
types are described in Section~\ref{sec4}. In each section,
the stars appear in the constellation order.
The individual measurements are listed in Tables~\ref{tab1} and
\ref{tab2}, while
Table~\ref{tab3} gives parameters of spectroscopic orbits derived
for 9 binary stars discussed in Section~\ref{sec2}. 
The broadening functions for selected phases of the 9
contact binaries are shown in Figure~\ref{fig1}, while radial 
velocity orbital solutions for these stars are shown in 
Figure~\ref{fig2}. 
We show the broadening functions for single stars or
binaries without orbital solutions in Figure~\ref{fig3}.
The conclusions of the paper are summarized in 
Section~\ref{summ}.

\section{CONTACT BINARY SYSTEMS}
\label{sec2}

\subsection{V1464 Aql}

The variable star V1464~Aql (HIP~97600, HD~187438)
was suggested to be a contact binary by
\citet{Duer97} (hereinafter D97) on the basis of the 
low-amplitude Hipparcos light curve. 
We found that the star is indeed a close binary
and our six  observations confirm that the
period is two times longer than the original Hipparcos 
period and is equal to 0.697822 days.
The SB1 orbit is shown in Figure~\ref{fig2} while
the orbital elements are given in Table~\ref{tab3},
together with the remaining binary stars. The RV data
are in Table~\ref{tab2}, together with the
oher stars showing single-lined spectra.

A search for the presence of the 
signature of the secondary star in the BF was unsuccessful. 
The visible component shows a relatively large 
broadening of its spectral lines 
with $V \sin i = 94 \pm 4$ km~s$^{-1}$
(Figure~\ref{fig1}). 
The star is bright, $V_{max} = 8.6$. The Tycho-2
color index $B-V=0.237$ suggests, with no reddening, the
spectral type of A8 -- F0, much later than given in HDH,
A2~V. The Hipparcos parallax of $7.16 \pm 1.26$ mas suggests
$M_V=2.9$ at the light maximum, which corresponds to
F1/2V rather than early A. We do not have sufficiently many
spectral standards to attempt our own classification, but
the general appearance of our spectra for V1464~Aql 
indeed supports an early F spectral type.

\subsection{EL Aqr}

EL~Aqr (HIP~117317) was the subject of a previous DDO study
\citep[DDO-5]{ddo5} where the orbital coverage was good,
but the reported final elements had a larger scatter than for
most of the DDO systems, probably because of 
typically large zenith distances and of a 
relative faintness of the system at $V_{max}=10.35$.
The $B-V$ given in DDO-5 was incorrect; the  
value of $B-V=0.47$ better agrees with the DDO spectral type
of F3V, but still suggests some amount of interstellar reddening.
The spectral type is not available in HDH. 
For more information about the system, please consult DDO-5.

Our 15 observations are concentrated in the first half of
the orbit. They confirm the DDO results,
but the $K_2$ semi-amplitude is significantly smaller. This 
may be an indication of an insufficient spectral
resolution, although the peaks in the BF's are quite well
separated (Figure~\ref{fig1}).
The primary eclipse prediction of $T_0$ from the DDO observations 
served well the new observations and was adopted here without
a change.

\subsection{V759 Cen}

The bright contact binary V759~Cen (HIP~69256, HD~123732)
was discovered by \citet{Bond70}. In spite of
its brightness of $V_{max} \simeq 7.45$, it has not
been much observed since then with only sporadic
photometric observations for eclipse timing.
The color $b-y=0.39$ and the spectral type F8 
\citep{Bond70} or G0V (HDH) and the period of
0.394 days suggest a typical contact binary. 

The binary was observed 6 times within our program and 
these were its first RV observations. 
The phase distribution of the observations 
was far from optimal so that the orbital elements must
be treated as preliminary. As can be seen in Figure~\ref{fig1}, 
the spectral resolution was insufficient for this binary which
is probably visible at a low inclination angle.

The Cracow database consulted in April 2006 provided 
an ephemeris used for our observations, as given in Table~\ref{tab3}. 
Our bootstrap
estimated errors are very large because of the insufficient
number of observations. The mass ratio is probably close to
$q \simeq 0.2$.

\subsection{SX Crv}

The very interesting and important contact binary 
SX~Crv (HIP~61825, HD~110139) with the
currently smallest known mass ratio of $q \simeq 0.07$ 
\citep[DDO-5]{ddo5}
was observed 8 times. The BF's show the faint peak of
the secondary component quite well 
although the DDO-5 elements are definitely
better established as they were based on 49 best observations
selected from among 96 available ones. As may be 
expected for a lower 
spectral resolution, the current observations give 
a smaller $K_2$, but the center of mass velocity
appears to be also different; the latter effect may be
due to the uneven phase distribution of the observations.

For more information about SX~Crv, please consult DDO-5. 
Note the incorrect value of $B-V$ in that paper 
which should be 0.44. The HDH spectral type is F7V.

\subsection{VZ Lib}

The spectroscopically triple system VZ~Lib (HIP 76050) 
was analyzed in \citet[DDO-4]{ddo4}. 
The current observations show 
all three components, but the lower spectral resolution
results in a much stronger merging of the primary and
tertiary peaks in the BF. As observed for other binaries,
the value of the semi-amplitude $K_2$ is again smaller
for the current observations.
For more information about VZ~Lib, please consult DDO-4. 

In DDO-4, a continuous change of the radial velocity of
the companion in four seasons was noted. The current
observations with the mean value $V_3 = -36.7 \pm 2.9$ at
JD~2,451,034 (see Table~\ref{tab2} for individual observations)
confirm the $V_3$ variability within the
combined span of 4 years very well.
The ``kink'' in $V_3$ visible in Figure~5 in DDO-4
is apparently real so that the orbital period of
the triple system is probably quite short, of the
order of a few years. It may be necessary to look for 
systematic changes in the center of mass data for the
binary VZ~Lib itself to confirm its motion. The fact
that no obvious changes of $V_0$ have been noted 
so far suggests that the third component is probably much
less massive than the binary.

\subsection{DE Oct}

DE Oct (HIP 100187, HD 191803) was observed spectroscopically
for the first time within this program. D97 had suggested that
this is a contact binary with the orbital
period twice as long as the Hipparcos 
discovery period, $2 \times P = 0.5555922$ days. With this
period, our 3 observations cannot properly define an
orbit and no estimates of element uncertainties
could be determined. However, we can exclude applicability to
our observations of the Hipparcos conjunction 
time at $T_0 = 2,448,500.157$; the new value of 
$T_0$ is given in Table~\ref{tab3}. The BF's are poorly resolved so
that the measured velocities, particularly of the secondary component,
are very tentative. 

DE Oct is a visual binary with the angular separation of 22.9 arcsec
at the position angle $129^\circ$ and the magnitude difference between
the visual components of 2.85 (WDS~20194-7608). The secondary 
was far enough not to be included in the spectrograph slit.

The Tycho-2 color index $B-V=0.319$ suggests a spectral type
F1/2V, while the HDH spectral type is A9IV. The star is relatively
bright, $V_{max} = 9.15$.

\subsection{MW Pav}

MW Pav (HIP 102508, HD 197070) was observed within our
program for the first time and was the best observed star of
this series with 18 observations defining a good radial 
velocity orbit.
The spectral signatures are well separated in the BF's,
although one must take into account the warning signs from
the other binaries that $K_2$ might be systematically
underestimated at the available resolution. We assumed the
value of the period from the Hipparcos results. 

MW~Pav is a well known southern contact binary with
$V_{max}=8.80$, $B-V=0.33$ (Tycho-2), the spectral type F3IV/V
(HDH) and a relatively long orbital period of 0.795 day.
It was discovered by \citet{egg68} and initially designated
as BV~894. A light curve solution was presented by
\citet{Lapas80}. The secondary eclipse seemed to be
total, so that evaluation of the mass ratio appeared to
be possible. However, $q_{phot}=0.122 \pm 0.003$,
disagrees with our spectroscopic determination, 
$q_{sp}=0.228 \pm 0.008$, even if we consider a possibility
of a probable systematic under-estimate of $K_2$ by
(at most) 10\%. Our spectroscopic observation should
permit a combined solution of the parameters of this
binary.

\subsection{BQ Phe}

BQ~Phe (HIP~2005, HD~2145) was suggested by D97 to be a contact
binary with the period twice longer than given by the
Hipparcos discovery observations, $2 \times P = 0.437$ days.
We confirm that BQ~Phe is a contact binary, but with only
4 observations our orbital solution is indicative rather
than definitive and the formal errors are very large. We assumed
both the $T_0$ and the double Hipparcos period
(see Table~\ref{tab3}). 

The star was a bit faint for this program, 
$V_{max} = 10.4$. Its spectral type
F3/5V (HDH) agrees with $B-V=0.51$ (Tycho-2). 

\subsection{GR Vir}

GR Vir (HIP 72138) was analyzed for radial velocity
variations by \citet[DDO-2]{ddo2} where a good orbital 
solution was presented. With only
5 new observations we can only say that we fully confirm the
DDO-2 solution. We assumed both the $T_0$ and the period 
from the DDO-2 results. 

For more information about GR~Vir, 
please consult DDO-2. As for other systems observed 
before, we see that our value of $K_2$ is slightly
lower than that observed at DDO.

\section{POSSIBLE BINARY MEMBERS OF MULTIPLE SYSTEMS}
\label{sec3}

\subsection{CL Cet}

CL Cet (HIP 2274, HD 2554) was suggested in D97 to be 
a contact binary with a period twice longer than the
Hipparcos discovery result, $2 \times P = 0.6216$ days.
The star has $V_{max}=9.9$ and the Tycho-2 color index
$B-V=0.313$; the latter agrees with the spectral type of F2V
(HDH).

Our spectroscopic observations do not have sufficient
resolution to analyze apparent changes in the single-peaked,
wide broadening function (Figure~\ref{fig3}). It is
possible that the binary signature is masked by
a relatively rapidly rotating companion
with $V \sin i = 135 \pm 8$ km~s$^{-1}$. The single peak
in the BF has the velocity $V = -18.9 \pm 1.2$ km~s$^{-1}$.
However, a significant 
shift by 10 km~s$^{-1}$ from the average was observed 
for the last of our four observations. The case for a complex 
blending of three components in this 
system is the weakest one among the three cases
discussed here; the star may be in fact a pulsating one.

\subsection{CE Hyi}

CE Hyi (HIP 7682, HD 10270) is an other case suggested to be
a contact binary by D97. Again, the orbital period suggested
was $2 \times P = 0.4408$ days.

The star is known as a visual double star WDS 01389-5835
(HU~1553) with the angular separation of 1.9 arcsec at 
the position angle of
$10^\circ$ and a small magnitude difference of only 0.24.
Our 3 observations show very clearly that the spectrum is
dominated by a slowly rotating companion, while the close,
low-inclination, contact binary is visible only 
in the base of the combined BF profile. Hipparcos and 
Tycho photometry of individual components shows that it is
the fainter star (B) which is the photometric variable
and thus the contact binary.

The comparable light contribution of both components
to the combined spectrum is 
visible in the BF where the sharp-lined star shows
the peak with $V \sin i$ which is un-measurably low, below 
the spectral resolution of our observations, while
the contact binary light is distributed in the velocity 
domain within $\pm 200$ km~s$^{-1}$ (Figure~\ref{fig3}). 
The radial velocity of the slowly rotating companion
is $V_3=9.00 \pm 0.33$ km~s$^{-1}$.

The observed $V_{max} \simeq 8.3$ is for the combined 
light of both visual components. 
The Tycho-2 catalog gives $V_A=9.08$, $V_B=9.29$ 
and $(B-V)_A=0.333$, $(B-V)_B=0.497$,
respectively. The Simbad database gives $B-V=0.49$ 
and F5V for CE~Hyi. The spectral type is after HDH.

\subsection{V1084 Sco}

V1084 Sco (HIP 86294, HD 159705) was suggested by D97 to be
a contact binary with the period twice the Hipparcos
period, $2 \times P = 0.3003$ days. 

We have only 3 observations which show that system is 
a complex one: It appears to be a quadruple system consisting 
of a detached binary giving two
sharp peaks in the BF (see Figure~\ref{fig3}), and of a slightly 
fainter contact binary responsible
for the short-period photometric variability. 
The contact binary -- because of the stronger line broadening --
is just barely detectable at the base of the BF. 
The radial velocities of the sharp-line binary
components (designated as ``3'' and ``4'' in Table~\ref{tab2})
varied during the 3 days of observations between $-19$ and $-31$
km~s$^{-1}$ for the stronger component and +77 and +83 
km~s$^{-1}$ for the fainter component. Thus, the detached
binary must be also relatively compact, but our observations were
insufficient to determine any parameters of the radial
velocity orbit.
The star was included in the major radial velocity survey of
\citet{Nord2004} where it appears with the average radial
velocity of +21.3 km~s$^{-1}$.

This star is a very interesting object for further studies,
particularly if the mutual period of revolution of the two
binaries turns out moderately short to be observable.
The star is relatively 
bright, $V_{max}= 9.0$, while the color and the spectral
type given in Simbad are late, $B-V=0.76$ and G6V (HDH).
The Tycho-2 catalog is in agreement with $B-V=0.73$.

\section{RADIAL VELOCITY VARIABLES OF UNKNOWN TYPE}
\label{sec4}

\subsection{V872 Ara}

This star, at that time identified as HIP~81650 (HD~149989)
was suspected in D97 to be a contact binary with the orbital
period of 0.8532 days. Very little can be said on the basis of its 
light variations which are very small (0.02 mag.). Three
observations obtained here show a wide, rotationally broadened
profile with the average $V \sin i = 142 \pm 6$ km~s$^{-1}$.
The mean velocity is constant at $+42.1 \pm 2.4$
km~s$^{-1}$, but the variation between +37 and +45 km~s$^{-1}$
is larger than the measurement error of about $\pm 1.2$
km~s$^{-1}$ so that some small variability may be present.

Our results are fully consistent with the recent 
study of \citet{deCat2006} which explains the variability of
V872~Ara by $\gamma$~Dor-type pulsations with the originally
suggested period of 0.42658 days. The measured value of
$V \sin i = 134 \pm 3$ km~s$^{-1}$ is consistent within
the combined errors with our estimate. We refer the reader
to the paper of \citet{deCat2006} for more information on this
star. The spectral type is A8/F0V (HDH).

\subsection{BD Cap}

BD~Cap (HIP~99365, HD~191301) was suggested by D97 to be
a contact binary with the period twice as long as the one
given by the Hipparcos project, $2 \times P = 0.3204$ days.
Our three spectra show a very broad BF with 
$V \sin i = 133 \pm 10$ km~s$^{-1}$. The mean velocity is
practically constant at $+9.7 \pm 1.0$ km~s$^{-1}$. We
cannot say more about this star except we note that
it was included in the catalog of suspected and confirmed
$\delta$~Sct pulsating stars \citep{Rod2000} as well as
in the survey of spatial velocities of nearby stars 
\citep{Nord2004}. The spectral type is A9III (HDH).

\subsection{Anon Cen = HIP~69300}

HIP~69300 (HD~123720) was an other suggestion of D97 to be a contact
binary. Our two observations substantially differ in radial
velocity of the star, $-94.6$ and $-25.9$ km~s$^{-1}$, but
the broadening profile has the same $V \sin i = 116 \pm 7$
km~s$^{-1}$. The star does not have an entry in the 
General Catalog of Variable 
Stars\footnote{http://www.sai.msu.su/groups/cluster/gcvs/,
the most recent electronic version 4.2.} and no variable
star name has been assigned to it yet, but it is definitely
a radial velocity variable. The spectral type is A4V (HDH).

\subsection{BX Ind}

BX Ind (HIP~108741, HD~208999), an other candidate of
D97, appears to be a slowly rotating star. Our seven 
observations all show a BF peak consistent with no
rotation. Some small radial velocity changes within
$-32$ and $-20$ km~s$^{-1}$ appear to be present with
the mean value $-27.6 \pm 1.7$ km~s$^{-1}$. This is
definitely not a close binary star. It is listed in the
Catalog of Delta Scuti stars of \citet{Rod2000}.
The spectral type is F2V (HDH).

\subsection{V388 Pav}

We have only two observations of 
V388 Pav (HIP~103803, HD~199434),
another candidate of D97. The radial velocity may be
constant at the mean of $+5.6 \pm 1.2$ km~s$^{-1}$, while
the BF's indicate a mild broadening of $V \sin i = 45 \pm 7$
km~s$^{-1}$. It is not a close binary star. It is listed in the
Catalog of Delta Scuti stars of \citet{Rod2000}.
The spectral type is F5II (HDH).

\section{SUMMARY}
\label{summ}

This program of radial velocity measurements of known and
suspected southern contact binary stars was performed to fill the
growing disparity in the available RV data for northern and southern
hemispheres. With only four successive nights, the program
could not achieve the same goals as the current David Dunlap 
Observatory survey. Still some useful results have been obtained
for 17 targets of the Fall southern sky.

We have confirmed the suggestion 
of \citet{Duer97} (D97) that V1464~Aql, 
DE~Oct, BQ~Phe are contact binaries and obtained the first 
preliminary orbital data for these systems; V1464~Aql
is a single-lined binary (SB1) while the rest are
double-lined systems. 
We obtained the first radial velocity orbital data (SB2) for
the well known southern systems V759~Cen and MW~Pav.
We confirmed the David Dunlap Observatory 
results for the double-lined binaries EL~Aqr, SX~Crv, VZ~Lib, GR~Vir,
but we noticed that in all these systems the secondary star
semi-amplitude $K_2$ is by a few percent smaller than 
observed at DDO which may be a
result of the lower spectral resolution. 

Three systems could not be analyzed because of the presence
of companions. In the case of CE~Hyi, a visual companion had 
been known, but we see spectral signatures of a binary companion
in V1084~Sco (so that the system is a quadruple one)
and suspect a presence of a companion in CL~Cet. 
We are not able to say much about other variables suggested in D97:
V872 Ara, BD~Cap, HIP~69300, BX~Ind, V388~Pav, but most appear
to be pulsating stars and have been included in 
catalogs of such objects; we give their mean radial velocities 
and $V \sin i$.

\acknowledgements

Thanks are due to George Conidis for participation in reductions
of the data used in this paper. Thanks are also due to the
reviewer, Dr.\ Vakhtang S.\ Tamazian, for a few very pointed
suggestions on the improvement of the paper.

Support from the Natural Sciences and Engineering Council of Canada
to SMR is acknowledged with gratitude. 
The research made use of the SIMBAD database, operated at the CDS, 
Strasbourg, France and accessible through the Canadian 
Astronomy Data Centre, which is operated by the Herzberg Institute of 
Astrophysics, National Research Council of Canada.
This research made also use of the Washington Double Star (WDS)
Catalog maintained at the U.S. Naval Observatory and the
General Catalog of Variable Stars maintained at the
Sternberg Astronomical Institute, Moscow, Russia.

\clearpage

\noindent
Captions to figures:

\bigskip

\figcaption[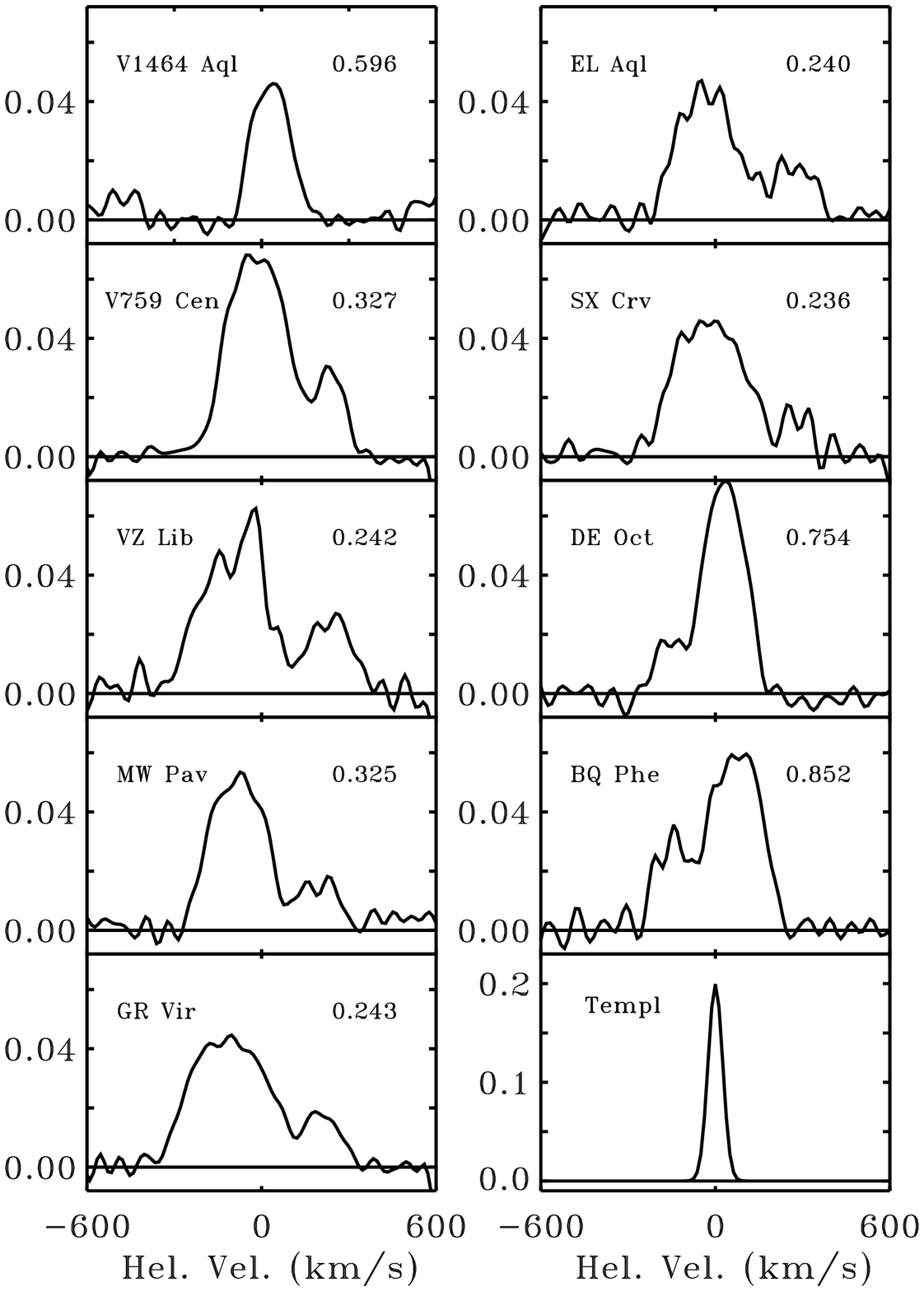]{\label{fig1} Broadening functions for 9 contact
binary systems discussed in Section~\ref{sec2}. The orbital phase is
given in the right side of each panel. The last panel gives the BF
representing the nominal resolution of the method.
}

\figcaption[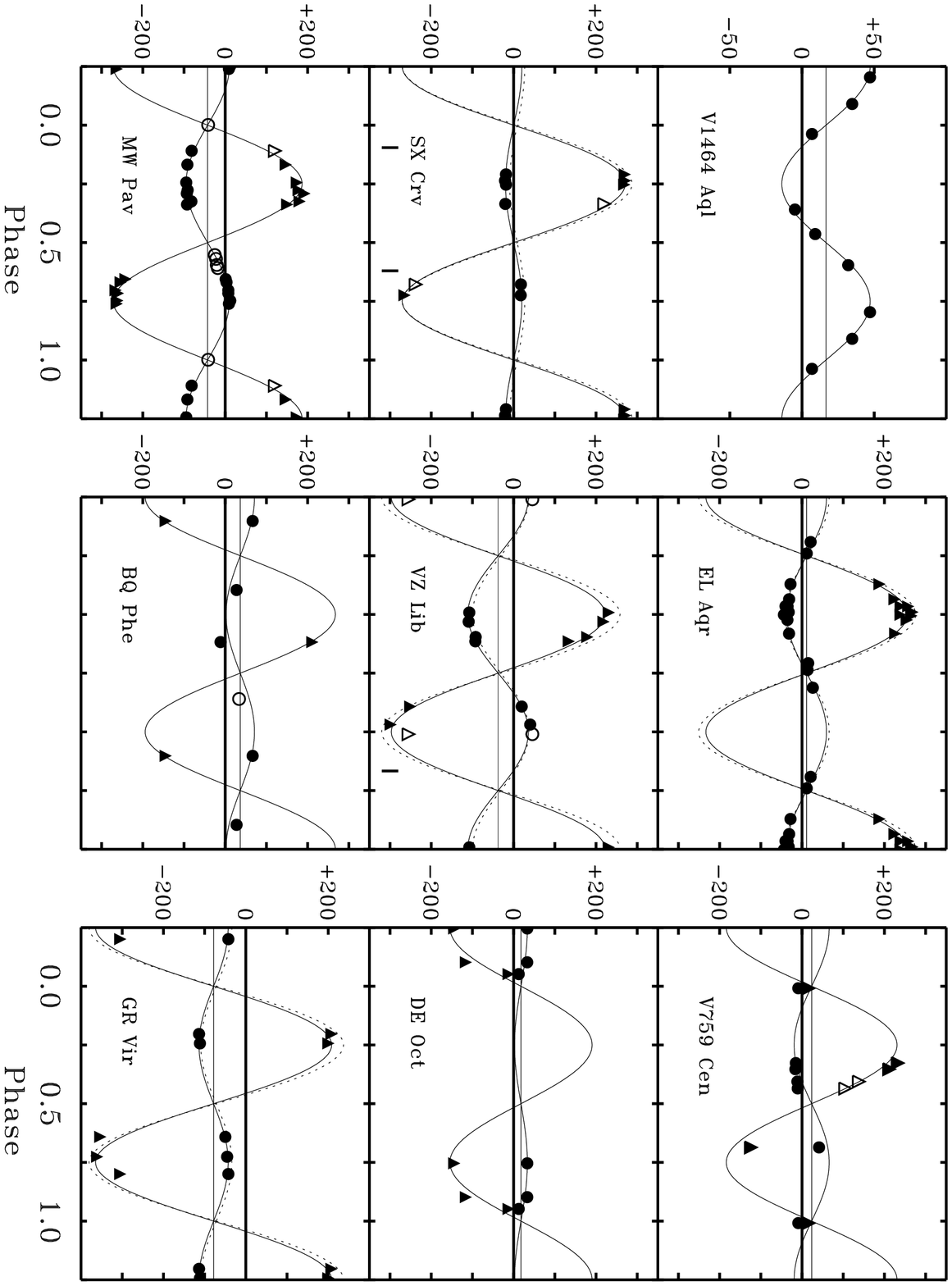]{\label{fig2} The orbital solutions for 9 
contact systems discussed in Section~\ref{sec2}. Observations of
lower quality are marked by open symbols. Dashes at the bottom mark
orbital phases when signatures of the components were unresolved.
The sine curve solutions based on DDO data are shown by broken lines.
Note that present solutions give systematically smaller values of
$K_2$. V1464~Aql is the only single lined binary (SB1) in this
group of stars.
}

\figcaption[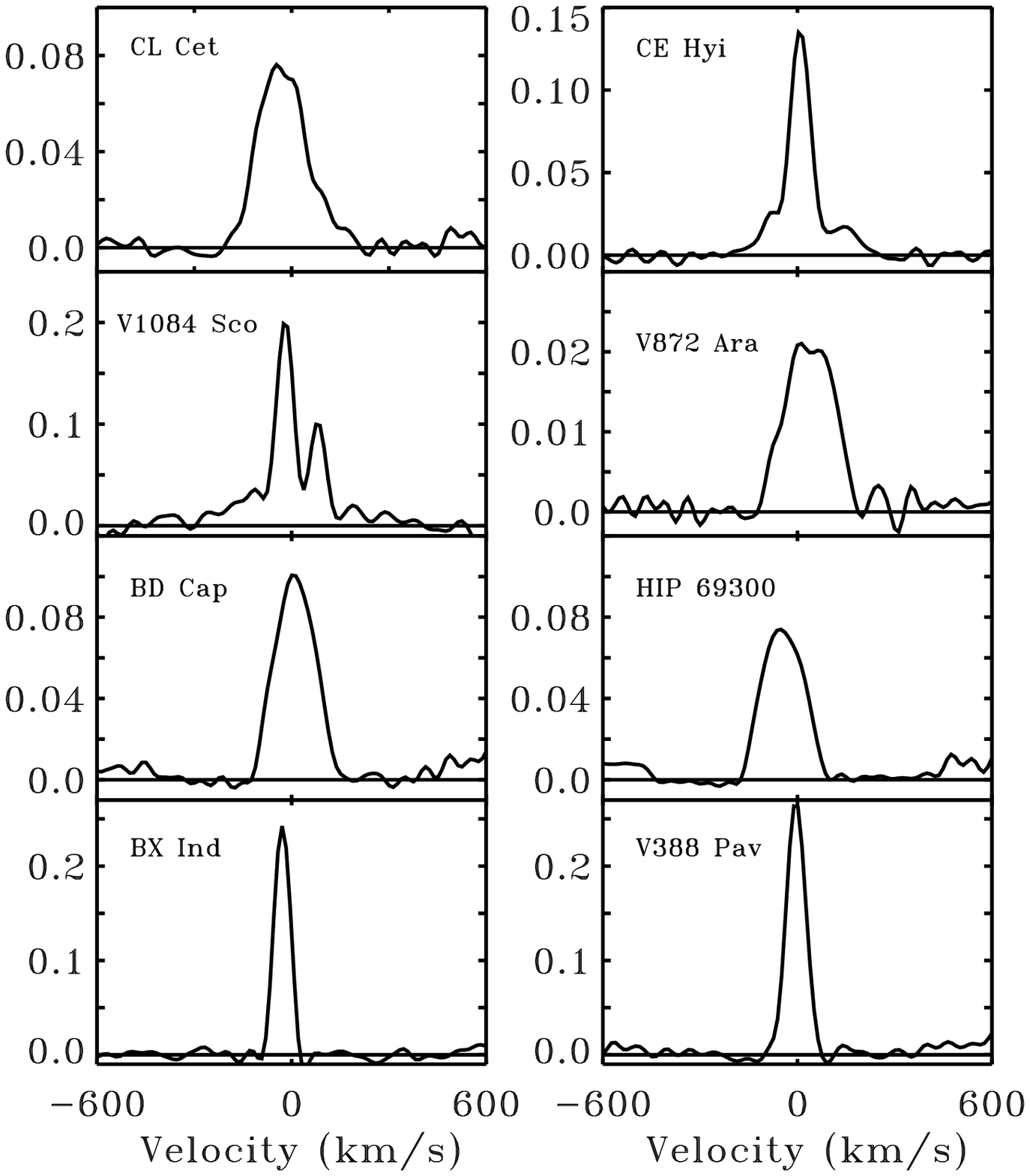]{\label{fig3} The 3 first panels
show broadening functions for multiple systems discussed 
in Section~\ref{sec3}, while the next 5 panels show the respective
functions for radial velocity variables of mostly unknown type, as
discussed in Section~\ref{sec4}. Note that ``strengths'' or
``intensities'' of the BF's have different vertical scales 
for different stars. This is because the BFs
depend not only on the geometric (rotational) broadening, but
also on how well spectral types of the template and of the
program star match. For a perfect fit, the integral of the BF should give
unity. The Y-axis units correspond to the BF sampling at
12.5 km~s$^{-1}$ per point.
}


\begin{deluxetable}{lrrrrr}


\tablewidth{0pt}
\tablenum{1}

\tablecaption{Radial velocity observations of double lined binaries
(the full table is available in electronic form)
\label{tab1}}
\tablehead{
\colhead{Star}     &
\colhead{HelJD}    & 
\colhead{~$RV_1$} & \colhead{~~$W_1$} &
\colhead{~$RV_2$} & \colhead{~~$W_2$} 
}
\startdata
EL Aqr  &    2451033.7239  &     20.84   &   1.0  &    0.00  &     0.0 \\
EL Aqr  &    2451033.7474  &     11.39   &   1.0  &    0.00  &     0.0 \\
EL Aqr  &    2451033.8103  &  $-27.96$   &   1.0  &  187.98  &     1.0 \\
EL Aqr  &    2451033.8556  &  $-40.11$   &   1.0  &  239.22  &     1.0 \\
EL Aqr  &    2451033.8732  &  $-43.90$   &   1.0  &  238.97  &     1.0 \\
\enddata

\tablecomments{The table gives the radial velocities $RV_i$ and
associated weights $W_i$ for observations of 8 stars described
in Section~\ref{sec2}. The velocities are expressed in km~s$^{-1}$. 
The weights $W_i$ were used in the orbital solutions and can take
values of 1.0, 0.5 or 0; the zero weight observations
may be eventually used in more extensive modeling of broadening
functions.}
\end{deluxetable}

\begin{deluxetable}{lrr}


\tablewidth{0pt}
\tablenum{2}

\tablecaption{Radial velocity observations of single lined 
stars (the full table is available in electronic form)
\label{tab2}}
\tablehead{
\colhead{Star}     &
\colhead{HelJD}    & 
\colhead{~~$RV$} 
}
\startdata
V1464 Aql &  2451033.6816  &    9.13  \\
V1464 Aql &  2451033.7740  &   32.03  \\
V1464 Aql &  2451034.7799  &    6.90  \\
V1464 Aql &  2451035.7022  &   $-4.92$   \\
V1464 Aql &  2451036.7050  &   47.12  \\
\enddata

\tablecomments{The table gives the radial velocities $RV$ 
for observations of stars described in Sections~\ref{sec2}
and \ref{sec3}. The velocities are expressed in km~s$^{-1}$. 
}
\end{deluxetable}

\begin{deluxetable}{lcccccccccccc}

\tabletypesize{\scriptsize}

\pagestyle{empty}

\tablecolumns{13}

\tablewidth{0pt}

\tablenum{3}
\tablecaption{Spectroscopic orbital elements for 9 contact binaries
\label{tab3}}
\tablehead{
   \colhead{Name} &                
   \colhead{DDO}  &                
   \colhead{n$_{obs}$}  &          
   \colhead{n$_{used}$} &          
   \colhead{$V_0$} &               
   \colhead{$\sigma V_0$} &        
   \colhead{$K_1$} &               
   \colhead{$\sigma K_1$} &        
   \colhead{$K_2$} &               
   \colhead{$\sigma K_2$} &        
   \colhead{$T_0$ -- 2,400,000} &  
   \colhead{$\sigma T_0$} &        
   \colhead{P}              
}
\startdata
V1464 Aql & new & 6 &  6 & $+16.61$ & 1.58 & 30.62 & 2.35 &        &      & 48,500.2642  & 0.0069 & [0.697822]\\[1.5ex]

EL Aqr    & 5   &41 & 31 & $+12.51$ & 2.08 & 53.38 & 1.53 & 263.34 & 4.38 & 51,109.3334  &  0.0032 & [0.481410]\\
          & new &15 & 14 & $+11.21$ & 2.90 & 48.12 & 3.13 & 244.57 & 5.63 &  [same]      & \nodata &  [same]\\[1.5ex]

V759 Cen  & new & 6 &  6 & $+23.4$ & 10.7  & 42.5 & 8.5 & 207.2 & 42.5 & [52,500.368] & \nodata & [0.3939999]\\[1.5ex]

SX Crv    & 5   &96 & 49 &  $+8.71$ & 0.94 & 18.28 & 0.74 & 278.70 & 2.43 & 51,070.8192  & 0.0010  & [0.316622]\\
          & new & 8 &  6 &  $+0.17$ & 1.35 & 19.79 & 1.24 & 270.60 & 1.96 &  [same]      & \nodata &  [same]\\[1.5ex]

VZ Lib    & 4   &61 & 38 & $-31.11$ & 2.30 & 68.54 & 3.84 & 289.25 & 4.55 & 51,091.4297  & 0.0015  & [0.358263]\\
          & new & 8 &  7 & $-37.64$ & 2.69 & 73.32 & 2.48 & 259.36 & 4.51 &  [same]      & \nodata &  [same]\\[1.5ex]

DE Oct    & new & 3 &  3 & $+17.86$ & 0.96 & 15.42 & 1.30 & 172.55 & 1.19 & 48,500.026   & 0.020   & [0.5555922]\\[1.5ex]

MW Pav    &new  &18 & 18 & $-42.75$ & 1.38 & 52.35 & 1.15 & 229.34 & 3.52 & 48,500.098   & 0.020   & [0.7949810]\\[1.5ex]

BQ Phe    & new & 4 &  4 & $+36.52$ & 8.6  & 34.44 & 5.6  & 230.9 & 80.1 & [48,500.085] & \nodata & [0.436968]\\[1.5ex]

GR Vir    & 2   &43 & 37 & $-71.72$ & 0.89 & 37.78 & 1.00 & 308.81 & 1.96 & 50,541.9582  & 0.0008  & [0.346979]\\
          & new & 5 &  5 & $-78.26$ & 1.39 & 35.39 & 2.27 & 286.55 & 2.39 &  [same]      & \nodata &  [same]\\[1.5ex]
\enddata
\tablecomments{The column DDO gives the number of the DDO paper
 or ``new'' for this paper. 
n$_{obs}$ and n$_{used}$ give the number of available 
and used RV measurements, respectively. 
The radial velocity parameters $V_0$, $K_1$ and $K_2$ and their rms 
errors are in km~s$^{-1}$.
$T_0$ is the heliocentric Julian Day of the superior conjunction 
(eclipse). The period $P$ is in days.
The assumed and fixed quantities are in square brackets. 
}
\end{deluxetable}


\newpage
\includegraphics{f1.eps}
\newpage
\includegraphics[angle=180]{f2.eps}
\newpage
\includegraphics{f3.eps}

\end{document}